\documentclass{article}
\usepackage[submission]{colm2026_conference}
\usepackage[utf8]{inputenc}
\usepackage[T1]{fontenc}
\usepackage[expansion=false]{microtype}
\usepackage{hyperref}
\usepackage{url}
\usepackage{booktabs}
\usepackage{amsmath,amssymb,amsthm,mathtools}
\usepackage{enumitem}
\usepackage{array}
\usepackage{tabularx}
\usepackage{xcolor}
\usepackage{graphicx}
\usepackage{tikz}
\usetikzlibrary{positioning,arrows.meta,fit,backgrounds,calc,decorations.pathreplacing}
\usepackage{algorithm}
\usepackage{algpseudocode}
\usepackage{pgfplots}
\pgfplotsset{compat=1.18}
\usepackage{pifont}

\definecolor{darkblue}{rgb}{0, 0, 0.5}
\hypersetup{colorlinks=true, citecolor=darkblue, linkcolor=darkblue, urlcolor=darkblue,
            hypertexnames=false, plainpages=false}

\newtheorem{definition}{Definition}[section]

\newtheorem{proposition}[definition]{Proposition}

\newcommand{\cometh}{\textbf{Comet-H}}

\newcommand{\W}{\mathcal{W}}

\newcommand{\Reals}{\mathbb{R}}

\newcommand{\AM}{\mathfrak{M}}
\newcommand{\ObVec}{\mathbf{o}}
\newcommand{\decay}{\lambda}
\newcommand{\pressure}{p}
\newcommand{\scorer}{\phi}
\newcommand{\athree}{{\sf a3}}
\newcommand{\papercomment}[1]{}

\newcommand{\pp}[1]{{{\mathsf{#1}}}}
\newcommand{\pIdeation}{\pp{Ideation}}
\newcommand{\pTheoryCreation}{\pp{TheoryCreation}}
\newcommand{\pSeedGeneration}{\pp{SeedGeneration}}
\newcommand{\pSeedUpgrade}{\pp{SeedUpgrade}}
\newcommand{\pPaperStrengthening}{\pp{PaperStrengthening}}
\newcommand{\pREADMEVerification}{\pp{READMEVerification}}
\newcommand{\pBenchmarkTightening}{\pp{BenchmarkTightening}}
\newcommand{\pGroundingCreation}{\pp{GroundingCreation}}
\newcommand{\pSkepticalAudit}{\pp{SkepticalAudit}}
\newcommand{\pPaperRewrite}{\pp{PaperRewrite}}
\newcommand{\pClaimCleanup}{\pp{ClaimCleanup}}
\newcommand{\pPortfolioExpansion}{\pp{PortfolioExpansion}}
\newcommand{\pFinalGroundingAudit}{\pp{FinalGroundingAudit}}

\newcommand{\pAcademicPaperPolish}{\pp{AcademicPaperPolish}}
\newcommand{\pBenchmarkSearch}{\pp{BenchmarkSearch}}

\title{Theory Under Construction: Orchestrating Language Models\\for Research Software Where the Specification Evolves}
\author{Halley Young\thanks{Microsoft Research} and Nikolaj Bj{\o}rner\thanks{Microsoft Research}}

\begin{document}

\maketitle

\begin{abstract}
  Large language models can now generate substantial code and draft research text.
  We are here aiming at a combination of code, empirical evaluation, and research text. In this pursuit, the
  hardest research-software projects fail while \emph{co-evolving} the connected artifacts:
  the mathematical thesis, the executable system, the benchmark surface,
  and the public claims must mature together, and they routinely drift apart.
  We identify two LM-specific failure modes---\emph{hallucination accumulation}, 
  where claims exceed what code or theory supports and unsupported assertions propagate across sessions,
  and \emph{desynchronization}, where code, theory, or the LM's own world model fall out of alignment with one another.

  We propose \emph{Comet-H}: an iterative prompt automaton that orchestrates
  ideation, implementation, evaluation, grounding, and paper-writing as coupled coordinates of a single workspace state.
  At each step the controller picks the next prompt by scoring it against what the workspace currently lacks,
  carries unfinished follow-up work forward with a half-life so old items fade unless they are addressed,
  and re-checks the paper and README against the underlying code and benchmarks whenever the documentation changes.
  We view this prompt-selection step as a small contextual bandit problem over prompt families---prompts are arms,
  workspace deficits are the context, and the score is a plain linear function whose weights are fixed by hand---and
  argue that pairing such a transparent scorer with a fading record of unfinished work is itself a useful design point:
  it keeps long-horizon follow-ups bounded, requires no learned policy, and makes every prompt choice easy to read off the workspace.

  We created a full portfolio spanning 46 research-software repositories across two dozen domains.
  We study \athree{} in depth: a Python static-analysis tool built entirely within the loop
  that reaches $F_1 = 0.768$ on a 90-case benchmark against a next-best baseline of $0.364$,
  with monotone ablation gains from each added theory/practice coupling layer.
  We report observations on LM behavior across ${\sim}400$ commits of orchestrated development,
  including the finding that audit-and-contraction passes dominate the later phases of every successful trajectory.
\end{abstract}

\section{Introduction}\label{sec:intro}

Language models are now capable code producers.
Systems like SWE-agent~\citep{yang2024sweagent},
Devin~\citep{cognition2024devin},
and OpenHands~\citep{wang2024openhands} can resolve GitHub issues, pass SWE-bench tasks, and generate multi-file repositories from specifications.
FunSearch~\citep{romera2024funsearch} and AlphaEvolve~\citep{novikov2025alphaevolve} show that LM-guided search can discover novel algorithms.
The AI Scientist~\citep{lu2024aiscientist} automates the full cycle from ideation through paper writing.
Mainstream coding agents Claude~\cite{claudecode}, Codex~\cite{codex}, Co-Pilot~\cite{copilot}, Gemini~\cite{googlecode}, Kiro~\cite{kiro}
are rapidly democratizing the capabilities of the earlier systems.
But they cannot be trusted naively---hallucinations, plausible garbage, and point solutions that address symptoms instead of root causes are the norm.
Much effort has been spent trying to ``ask the right questions'' to LMs to avoid bad outcomes.

Our approach is different: instead of crafting better prompts, we \emph{design a process architecture} that (i)~determines \emph{when} to ask \emph{what}, (ii)~determines when to trust or test the answers, (iii)~allows goals, claims, theory, code, and evidence to evolve together, and (iv)~scales useful compute trajectories to days, not minutes.

\begin{figure*}[t]
\centering
\includegraphics[width=0.98\textwidth]{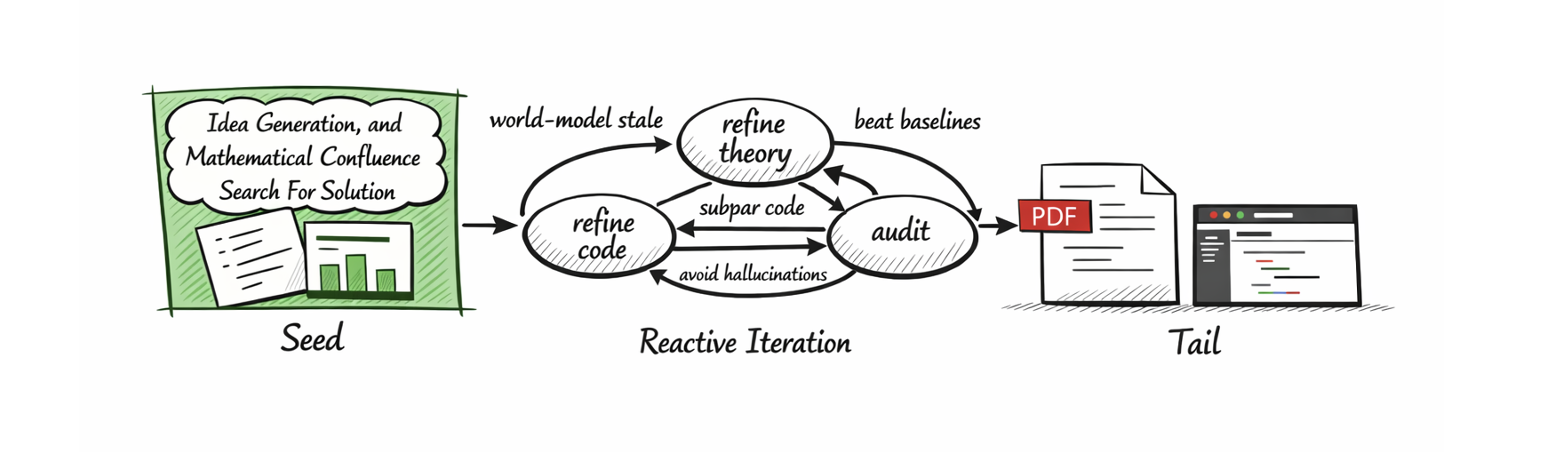}
\caption{High-level \cometh{} loop: the system cycles through ideation, generation, hardening, and audit until a batch of repositories reaches quality thresholds. The  controller that implements this loop is described in Section~\ref{sec:formalism}.}
\label{fig:loop}
\end{figure*}

The hardest research-software projects do not fail at generation.
They fail based on drift: when the mathematical thesis, the executable system, the benchmark surface, and the public claims drift apart.
When these coupling edges are ignored, two characteristic failure modes emerge:
\emph{hallucination accumulation}, where claims exceed what the code or theory actually supports and unsupported assertions propagate;
and \emph{desynchronization}, where code or theory advances past the claims, or where even pairwise consistency between theory, code, and claims is insufficient to guarantee global coherence.
This \emph{co-evolution problem} is invisible to existing evaluation frameworks because those frameworks assume a fixed external task.
Our \cometh{} process creates research artifacts---code and papers---using a state machine transitioning along states
illustrated in Figure~\ref{fig:loop}. It starts with an initial seed, exploring a research direction; then through reactive iteration
it produces research artifacts with the \cometh{} process ensuring that theory co-evolves with empirical feedback.

Research software development occupies a regime (Table~\ref{tab:distinction}) that requires both mathematical ambition and repository scale, but where the specification itself is under construction.
Prior systems either assume a fixed evaluation frame (FunSearch~\citep{romera2024funsearch}, AlphaEvolve~\citep{novikov2025alphaevolve}), operate on shorter
horizons with fixed tasks (SWE-agent~\citep{yang2024sweagent}, Devin~\citep{cognition2024devin}),
or automate ideation without a durable repository loop (AI Scientist~\citep{lu2024aiscientist}).
Recent long-horizon systems such as Ralph-style persistent engineering loops~\citep{huntley2026ralph} and AutoResearchClaw-style staged research pipelines~\citep{liu2026autoresearchclaw} move closer to this regime, but still center engineering closure or paper-production pipelines rather than explicit control of cross-artifact co-evolution.
None addresses the full co-evolution problem.

\begin{figure*}[t]
\centering
\begin{tikzpicture}[
  box/.style={draw, rounded corners=5pt, fill=#1,
              minimum width=4.8cm, minimum height=1.0cm,
              align=center, inner sep=7pt, font=\small},
  side/.style={draw, rounded corners=5pt, fill=#1,
               minimum width=3.2cm, align=center, inner sep=7pt, font=\small},
  arr/.style={-{Stealth[length=6pt]}, thick},
  oarr/.style={-{Stealth[length=5pt]}, semithick, red!55!black},
  tarr/.style={-{Stealth[length=5pt]}, semithick, blue!55!black},
]

\node[box={blue!18}] (ws) at (0,5.5)
  {Workspace $W_t$\\[-2pt]{\scriptsize theory, code, paper, evidence}};

\node[box={yellow!20}] (adm) at (0,3.7)
  {Kernel $\kappa$: admissible set $P_t$\\[-2pt]{\scriptsize mode guards $\cap$ forced queue $\cap$ budget}};

\node[box={green!18}] (score) at (0,1.9)
  {Scorer $\scorer$\\[-2pt]{\scriptsize select $p_t \in \arg\max_{p \in P_t} \scorer(W_t,\ObVec_t)_p$}};

\node[box={violet!12}] (exec) at (0,0.1)
  {Update $\delta$: execute prompt $p_t$\\[-2pt]{\scriptsize LM run edits repository; $(W_{t+1}, m_{t+1}) = \delta(W_t, m_t, p_t)$}};

\draw[arr] (ws)    -- (adm)    node[midway, right, font=\small] {workspace deficits};
\draw[arr] (adm)   -- (score);
\draw[arr] (score) -- (exec);
\draw[arr, rounded corners=6pt]
  (exec.west) -- ++(-1.0,0) |- (ws.west)
  node[pos=0.42, left, font=\small] {$W_{t+1}$ re-read from disk};

\node[side={pink!20}, minimum height=1.2cm] (trig) at (6.2,2.8)
  {Grounding trigger $\tau$\\[-2pt]
   {\small paper/README $\Delta \Rightarrow$ force audit}};

\draw[tarr, rounded corners=4pt]
  (exec.east) -| (trig.south)
  node[pos=0.25, above, font=\small] {diff};
\draw[tarr, rounded corners=4pt]
  (trig.west) -| (adm.south);

\end{tikzpicture}
\caption{\cometh{} controller architecture.
  Workspace $W_t$ feeds selection kernel~$\kappa$ (mode guards, forced queue, budget).
  Scorer~$\scorer$ selects~$p_t$; update map~$\delta$ executes the prompt and re-reads the repository.
  Paper or README changes trigger a forced grounding-then-audit pass (Proposition~\ref{prop:trigger}).}
\label{fig:architecture}
\end{figure*}
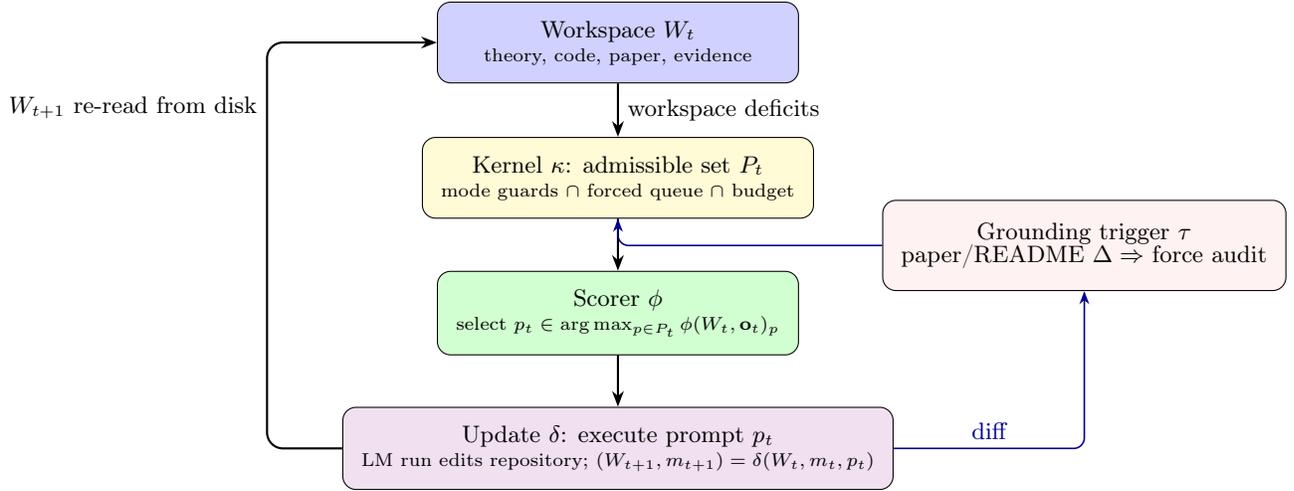

This paper studies \cometh, an iterative prompt automaton for this regime (detailed in \S\ref{sec:process}--\S\ref{sec:formalism}; see Figure~\ref{fig:architecture}).
The standard pipeline generates batches of ten repositories per run; across multiple runs, it has produced 46 research-software repositories spanning 12+ domains.
Each repository contains source code, a tool paper, a machine-readable grounding ledger, benchmark harnesses, and a README.
\footnote{
The experiments leading up to the \cometh{} process, and the subsequent
creation process for the portfolio involved running hundreds of sub-agents over several days.
It currently requires a \emph{very generous} budget for token consumption.
}
The key design principles are:

\begin{itemize} 
\item \textbf{Theory is mutable.} The underlying mathematical thesis of the research software
  is a state variable subject to revision under empirical pressure.
  This makes sense because the right formalization is rarely known at the outset: implementation reveals gaps, benchmarks expose false assumptions, and the correct abstraction only becomes clear once evidence accumulates.
\item \textbf{Artifacts and prompts are coupled.} Every prompt stage is keyed to the artifact surfaces it touches, ensuring no surface drifts unattended.
  Because prompts are generated from and evaluated against live artifacts, the system cannot silently diverge from what has actually been built.
\item \textbf{Artifact changes are audited.}
  Public-facing artifact changes force an immediate grounding-and-audit pass, bounding hallucination propagation to one step.
  Without this check, a single erroneous claim in a paper or README can cascade into downstream code changes and benchmark modifications before any inconsistency is detected.
\item \textbf{Drift is controlled by adjacency.}
  The research question may evolve, but only through adjacent moves that preserve the current evaluation surface.
  Unconstrained drift would allow the system to silently redefine the problem, rendering prior benchmarks irrelevant and making longitudinal comparison impossible.
\end{itemize}

\begin{table}[t]
\centering
\small
\caption{Comparison of \cometh{} against adjacent LM-based construction settings.
  \cometh{} is the only system where the metric, specification, and evaluation surface are all mutable.}
\label{tab:distinction}
\begin{tabularx}{\linewidth}{@{}>{\raggedright\arraybackslash}p{2.0cm}>{\raggedright\arraybackslash}p{2.0cm}>{\raggedright\arraybackslash}p{2.0cm}>{\raggedright\arraybackslash}p{1.0cm}>{\raggedright\arraybackslash}p{1.0cm}X@{}}
\toprule
\textbf{System} & \textbf{Scale} & \textbf{Duration} & \textbf{Metric fixed?} & \textbf{Multi-session?} & \textbf{Governance} \\
\midrule
FunSearch & function & mins & yes & no & none \\
AlphaEvolve & patch & hours & yes & no & none \\
AI Scientist & paper+code & 4--8\,h & yes & no & none \\
SWE-agent & patch & mins & yes & no & none \\
ChatDev & prototype & mins & yes & no & none \\
\cometh{} & \textbf{repo} & \textbf{24--48\,h} & \textbf{no} & \textbf{yes} & \textbf{first-class} \\
\bottomrule
\end{tabularx}
\end{table}

We make the following main contributions:
\begin{enumerate}
  \item 
    \textbf{A definition of the co-evolution problem and two LM-specific failure modes that cause it} (\S\ref{sec:problem}).
    This framing matters because existing benchmarks and agent evaluations assume a fixed, externally-supplied specification.
    In research software, the specification is itself a primary output: the mathematical thesis, the claimed scope, and the evaluation surface all change as the work matures.
    Ignoring this produces systems that optimize for local correctness while accumulating global incoherence across theory, code, and evidence.
  \item 
    \textbf{A Co-Evolution State Machine} (\S\ref{sec:process},\S\ref{sec:formalism}).
    The key insight is that a long development process can require many kinds of intervention — but the right intervention at any step is not arbitrary: it is inferrable from the current workspace state, and the interventions have systematic precedence relationships to one another.
    Formalizing this enables full automation across multi-session runs without manual steering.
  \item
    \textbf{A linear prompt scorer paired with a fading record of unfinished work} (\S\ref{sec:formalism}).
    Each step picks the next prompt by computing a small dot product between hand-set weights and a feature vector that summarises
    what the workspace currently lacks; pieces of unfinished follow-up work (e.g.\ ``benchmark this'', ``re-check this claim'')
    are stored separately and decay exponentially, so recent items dominate the score while stale ones drop out.
    We view this as a small contextual bandit over prompt families~\citep{li2010linucb,lattimore2020bandit}---one in which the
    underlying ``arm rewards'' shift across the run because the prompts themselves rewrite the workspace, much like in
    discounted-reward or sliding-window bandits for non-stationary environments~\citep{garivier2011discounted,besbes2014nonstationary}.
    The point of pairing a transparent linear scorer with a fading follow-up record is not to maximise asymptotic regret;
    it is to keep the orchestrator's choices auditable, bounded, and tied directly to readable workspace state, without any learned parameters.
  \item 
    \textbf{An empirical evaluation producing a portfolio of 46 research repositories using our methodology} (\S\ref{sec:a3},\S\ref{sec:evaluation}).
    To go in depth, we examine a flagship repository \athree{}. \athree{} is an extended bug-finding tool for Python that
    proves individual warnings safe by constructing small mathematical safety arguments and asking an SMT solver to confirm them;
    the underlying theory (an adaptation of \emph{barrier certificates} from control, originally used to prove that an unsafe state is unreachable) was selected and refined entirely within the loop. We provide external benchmarks, ablations, and real-codebase evaluation.
\end{enumerate}

\section{The co-evolution problem}\label{sec:problem}

Research software development differs from conventional software engineering in that the \emph{specification itself is under construction}.
The challenge when using LMs to evolve specifications is to ensure that they arrive at high quality goal states and don't devolve into local minima.

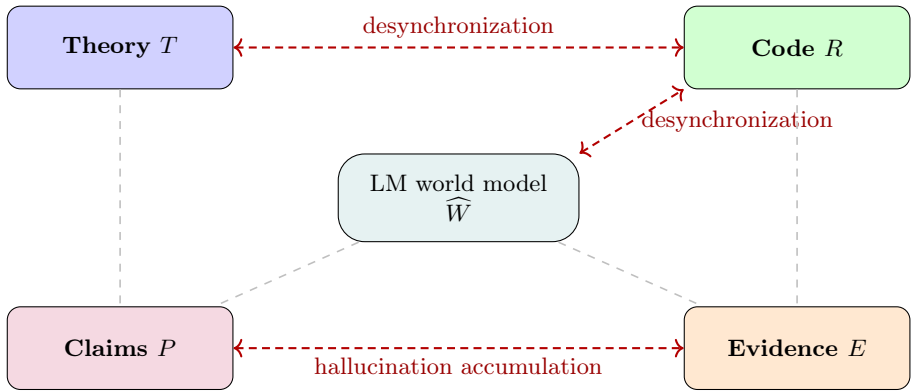
\begin{figure*}[t]
\centering
\begin{tikzpicture}[
  surf/.style={draw, rounded corners=5pt, fill=#1,
               minimum width=3.0cm, minimum height=1.1cm,
               align=center, inner sep=8pt, font=\small\bfseries},
  model/.style={draw, rounded corners=10pt, fill=teal!10,
               minimum width=3.2cm, minimum height=1.1cm,
               align=center, inner sep=8pt, font=\small},
  fm/.style={<->, red!70!black, thick, densely dashed},
  link/.style={black!25, semithick, dashed},
]
  \node[surf={blue!18}]   (theory)   at (0,0)     {Theory $T$};
  \node[surf={green!18}]  (code)     at (9.0,0)   {Code $R$};
  \node[surf={purple!15}] (claims)   at (0,-4.0)  {Claims $P$};
  \node[surf={orange!18}] (evidence) at (9.0,-4.0) {Evidence $E$};
  \node[model]            (world)    at (4.5,-2.0) {LM world model\\$\widehat{W}$};

  \draw[fm] (theory) -- (code)
    node[midway, above, font=\small, red!60!black] {desynchronization};
  \draw[fm] (claims) -- (evidence)
    node[midway, below, font=\small, red!60!black] {hallucination accumulation};
  \draw[fm] (world.north east) -- (code.south west)
    node[midway, right, font=\small, red!60!black, align=left] {desynchronization};

  \draw[link] (theory)  -- (claims);
  \draw[link] (code)    -- (evidence);
  \draw[link] (world)   -- (claims);
  \draw[link] (world)   -- (evidence);
\end{tikzpicture}
\caption{The two co-evolution failure modes, located at the artifact interfaces they destabilize.
  \textbf{Hallucination accumulation}: fabricated claims propagate through the loop because the LM cannot distinguish its own prior outputs from verified facts.
  \textbf{Desynchronization}: theory and code diverge---either because the LM optimizes code without revising the formal specification, or because its internal repository model goes stale as files accumulate beyond its context window.
  Dashed black lines show artifact dependencies that \cometh{} couples via cross-surface prompts.}
\label{fig:drift}
\end{figure*}

We frame the core difficulty as a graph of coupled artifact surfaces---theory, code, claims, and evidence---connected by edges that must stay synchronized.
When these coupling edges are ignored, two characteristic failure modes emerge: \emph{hallucination accumulation} and \emph{desynchronization}.
Figure~\ref{fig:drift} locates each failure mode at the coupling edges it destabilizes.

\paragraph{Hallucination accumulation.}
Claims exceed what the code or theory actually supports, and unsupported assertions propagate.
An LM rewrites a paper draft, introduces a fabricated number, and subsequent steps---which receive the paper as context---treat that number as ground truth.
Unlike single-turn hallucination (a well-studied problem), this is \emph{multi-step hallucination propagation}: the fabrication compounds through the development loop because the LM has no mechanism to distinguish its own prior outputs from verified facts.

\paragraph{Desynchronization.}
Code or theory advances past the other, or past the claims; sometimes, even pairwise consistency between theory, code, and claims is insufficient to guarantee global coherence.
Desynchronization has two faces.
First, \emph{world-model staleness}: after a long multi-agent execution that modifies dozens of files, the LM's representation of the repository is stale; the next prompt operates on an implicit world model that may not match disk state.
Second, \emph{specification drift}: the LM is excellent at optimizing code for a given objective, but in research software the objective itself should evolve.
Without explicit controller support, the model will keep improving code under the original specification even when implementation reveals that the formal framework should change.
This second face is the opposite of the usual ``drift'' concern: the problem is \emph{insufficient} drift, not excessive drift.

\paragraph{Addressing drift.}
Existing frameworks are not well equipped to deal with these failure modes.
Repository-scale agents (SWE-bench, issue resolution) assume a fixed task and evaluate against a fixed test suite---they have no mechanism for revising the task itself or auditing their own public claims.
Automated research systems (AI Scientist, SciAgents~\citep{lu2024aiscientist,SciAgents}) automate ideation and paper drafting but stop short of a durable loop in which claims must survive runnable code, benchmark harnesses, and typed grounding artifacts.
Neither family addresses the full co-evolution problem (Figure~\ref{fig:drift}).

\cometh{} is designed around these two failure modes rather than treating them as incidental errors.
It addresses hallucination accumulation through reactive grounding and skeptical audit passes that force public claims back into contact with executable evidence after documentation changes.
It addresses desynchronization by (i)~repeatedly re-reading and re-scoring the on-disk workspace state, so prompt selection is driven by current repository deficits rather than a stale internal summary, and (ii)~making theory revision an admissible controller action: adjacency-constrained expansion allows the formal objective to move when implementation and evaluation reveal that the initial specification is the wrong one.

\paragraph{Theory both should and will drift}
Our claim that ``in (automated) development, theory is mutable'' is both prescriptive and descriptive.  
It's descriptive in the sense that, unless clever measures are taken to the contrary, the theory a workflow implements will not be the theory that was first imagined - LLM's simply are not self-consistent enough for that.
It's prescriptive in the sense that most high-quality software research evolves as it is being produced - scientists realize that a hypothesis is incorrect or that an approach isn't promising and pivot, or discover and interesting new phenomenon and add that to the research agenda, displacing something else.  

\section{The \cometh{} process}\label{sec:process}

\begin{table}[t]
\centering
\small
\caption{Problems and Solutions in Automating Research Software}
\label{tab:lessons}
\begin{tabularx}{\linewidth}{@{}>{\raggedright\arraybackslash}p{3.2cm}>{\raggedright\arraybackslash}X@{}>{\raggedright\arraybackslash}X@{}}
\toprule
\textbf{Problem} & & \textbf{Solution in \cometh{}} \\
\midrule
\textbf{Desynchronization:} & Theory pivoted mid-build; no way to repurpose existing code &
Theory as explicit mutable state; mutations trigger re-hardening \\[4pt]
\textbf{Hallucination accumulation:} & LM was misled to assume successful code implementation of its theory existed &
Reactive grounding and auditing after every public-facing diff \\
\bottomrule
\end{tabularx}
\end{table}

At a high level, each run follows a sequence of reactive iterations (Figure~\ref{fig:loop}), implemented by a state machine that transitions over \emph{workspace} states.
A workspace captures the full context of a research-software project under construction: the mathematical theory being developed, the code repositories that implement it, the public-facing claims about it, the empirical evidence supporting those claims, the working hypothesis about what the software is useful for, and the outstanding quality debts that still need to be addressed.
We define these components precisely below, then describe how the controller selects prompts to evolve them.

\subsection{Workspace state and prompt families}

The controller operates over a workspace $W = (T, R, P, E, U, Q)$ with six components:
\begin{itemize}[nosep]
  \item $T$: the \textbf{theory}---the mathematical thesis, definitions, and formal core under development.  Without an explicit, mutable theory artifact, the mathematical content of the project exists only inside the LM's context window and drifts silently between invocations.
    Making theory a first-class workspace component forces every code change to stay anchored to a formal specification, and every specification change to propagate back into code---closing the theory--implementation desynchronization loop that otherwise produces repositories whose papers describe mathematics the code does not actually implement.
  \item $R$: the \textbf{repository forest}---the set of code repositories implementing and evaluating the theory.  The repository is the primary deliverable: installable, testable software.
    Tracking it as an explicit component lets the controller measure concrete progress (lines of implementation code, test coverage, installability) rather than relying on the LM's self-report, which we found to be unreliable.
  \item $P$: the \textbf{public projection}---The public projection is what reviewers and users actually see.
    Separating it from the repository lets the controller detect when claims drift from reality---the paper says ``3$\times$ faster'' but no benchmark supports it---and trigger corrective action before the drift compounds.
  \item $E$: the \textbf{evidence surface}---Evidence is the bridge between what the code does ($R$) and what the paper claims ($P$).
    Without an explicit evidence surface, the system has no way to distinguish grounded claims from hallucinated ones.
    The grounding ledger provides a machine-auditable record: every empirical number in the paper traces back to a runnable command and its output.
  \item $U$: the \textbf{utility hypothesis}---Research software must be \emph{useful}, not just correct.
    The utility hypothesis captures the ``why should anyone care?''~question and evolves as benchmarks reveal which niches the tool actually excels in.
    Without it, the system optimizes for mathematical elegance or code volume without regard to whether anyone would use the result; and
  \item $Q$: the \textbf{open obligations}---Obligations are the controller's memory of promises not yet kept.
    When a generation prompt creates new code and new paper claims, the obligation queue records that those claims have not yet been grounded or audited.
    This prevents the system from racing ahead with more expansion while a growing backlog of unverified claims accumulates---the primary mechanism by which hallucination cascades form.
\end{itemize}
At each step the controller re-reads the repository from disk to form the current workspace summary~$W_t$, then uses that summary to decide which of 17 fixed prompts to apply next.
This is a deliberate anti-desynchronization measure: rather than relying on the LM's internal memory of what the repository contains, the controller grounds every prompt-selection decision in the actual on-disk state, ensuring that the scored deficits reflect reality rather than a stale world model.

The prompt alphabet $\Sigma$ is organized into four phases (Table~\ref{tab:prompt-families}): seed (problem selection), generation (repository construction), hardening (iterative quality improvement), and tail (final audit).
In brief: \emph{seed} prompts define the research question; \emph{generation} prompts build the first runnable repositories; \emph{hardening} prompts iteratively strengthen theory, code, benchmarks, and claims; and \emph{tail} prompts finalize grounding and polish.
Each prompt is \emph{coded by the artifact surfaces it touches}---for example, a generation prompt that creates code ($R$) and a paper skeleton ($P$) is tagged as touching both surfaces, which forces the controller to schedule follow-up grounding on the evidence surface ($E$) before those new claims can propagate unchecked.
This coding couples artifact evolution by forcing cross-surface synchronization.

\begin{table}[t]
\centering
\small
\caption{Prompt families. \textbf{Phase}: S = seed, G = generation, H = hardening, T = tail. \textbf{Exp.}: expansive prompts that create follow-up grounding obligations.}
\label{tab:prompt-families}
\begin{tabularx}{\linewidth}{@{}>{\raggedright\arraybackslash}X>{\centering\arraybackslash}p{0.8cm}>{\centering\arraybackslash}p{0.8cm}>{\raggedright\arraybackslash}X@{}}
\toprule
\textbf{Prompt} & \textbf{Phase} & \textbf{Exp.} & \textbf{Role} \\
\midrule
Ideation ($\pIdeation$) & S & & differentiated thesis and formal core \\
Theory creation ($\pTheoryCreation$) & S & & mathematically substantive regime \\
Seed generation ($\pSeedGeneration$) & G & \checkmark & installable repos with paper skeleton \\
Seed upgrade ($\pSeedUpgrade$) & G & \checkmark & upgrade to serious grounded systems \\
Paper strengthening ($\pPaperStrengthening$) & H & & strengthen paper claims and theory \\
README verification ($\pREADMEVerification$) & H & & synchronize public examples with code \\
Benchmark tightening ($\pBenchmarkTightening$) & H & & strengthen measurement surfaces \\
Grounding creation ($\pGroundingCreation$) & H & & grounding ledger, contract claims \\
Skeptical audit ($\pSkepticalAudit$) & H & & audit claims against evidence \\
Paper rewrite ($\pPaperRewrite$) & H & & rewrite narrative for coherence \\
Claim cleanup ($\pClaimCleanup$) & H & & remove unsupported assertions \\
Portfolio expansion ($\pPortfolioExpansion$) & H & \checkmark & locally justified capability growth \\
Tail sequence ($\pFinalGroundingAudit$--$\pAcademicPaperPolish$) & T & & final grounding, critique, polish \\
Benchmark search ($\pBenchmarkSearch$) & H & & find real evaluation data \\
\bottomrule
\end{tabularx}
\end{table}

\subsection{Development loop}

A run produces a batch of ten repositories through three phases (Figure~\ref{fig:loop}): seed, reactive iteration (encompassing generation and hardening prompts), and tail.

\paragraph{Seed.} Stages $\pIdeation$--$\pTheoryCreation$ select differentiated problem theses and mathematical regimes.
These are part of research itself: the software is not well posed until a formal core has been selected.
\cometh{} treats the problem statement as part of the output.
The seed phase exists because, without explicit problem selection, the LM defaults to well-known tasks (e.g., ``build a transformer'') where it cannot plausibly beat established baselines.
Forcing ideation and theory creation up front ensures each repository targets a novel mathematical niche where a fresh implementation has a realistic chance of being competitive.

\paragraph{Reactive iteration.} The scorer examines workspace deficits and selects the highest-scoring admissible prompt.
Whenever the chosen prompt is one that creates new material---generating a fresh repository, upgrading a seed into a serious project, or expanding the portfolio---the controller automatically schedules the next two steps: first a grounding step that records what the new claims rest on, and then an audit step that checks them. This is what prevents unsupported assertions from accumulating across sessions.
The scoring responds reactively to workspace state: if benchmarks are weakest, benchmark-tightening fires; if the paper has drifted from code, paper-rewrite fires.
This reactive design is motivated by the observation that a fixed prompt order (e.g., always ``code then paper then benchmark'') wastes budget on surfaces that are already strong while neglecting surfaces that are falling behind.
The dot-product scorer continuously re-evaluates which surface is weakest and directs attention there, approximating the allocation strategy a human researcher would follow.

\paragraph{Tail.} A fixed nine-step sequence guarantees audit and paper coverage before budget exhaustion.
The tail exists because reactive scoring alone cannot guarantee that final-polish work (critique, response-to-critique, academic paper polish) will fire before the step budget runs out---the scorer might keep choosing code expansion if LOC gaps remain.
The tail overrides the scorer in the last steps to ensure every repository ends with a complete audit trail and a polished paper, regardless of what the scorer would have preferred.

\subsection{Reactive grounding triggers}

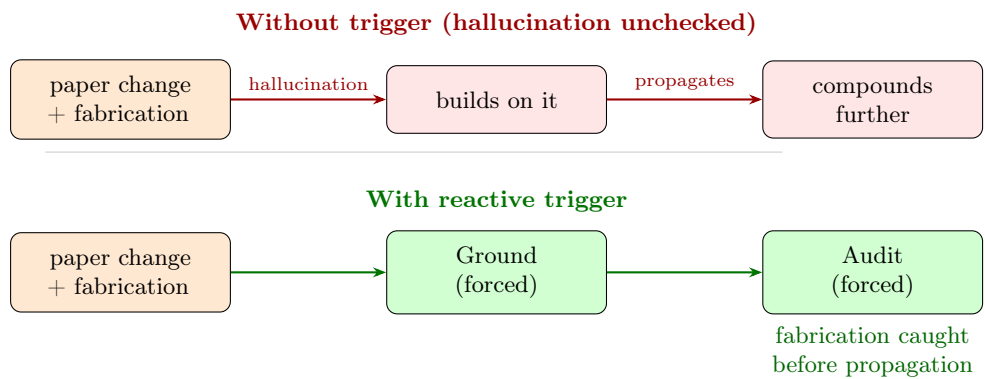
\begin{figure}[t]
\centering
\begin{tikzpicture}[
  box/.style={draw, rounded corners=4pt, fill=#1,
              minimum width=2.4cm, minimum height=0.9cm,
              text width=2.5cm, align=center, inner sep=6pt, font=\small},
  bad/.style={-{Stealth[length=5pt]}, thick, red!60!black},
  good/.style={-{Stealth[length=5pt]}, thick, green!45!black},
]
  \node[font=\small\bfseries, red!60!black] at (5.0,2.1) {Without trigger (hallucination unchecked)};
  \node[box={orange!18}] (p1) at (0,1.1)    {paper change\\+ fabrication};
  \node[box={red!10}]    (p2) at (5.0,1.1)  {builds on it};
  \node[box={red!10}]    (p3) at (10.0,1.1)  {compounds further};
  \draw[bad] (p1) -- (p2) node[midway, above, font=\scriptsize] {hallucination};
  \draw[bad] (p2) -- (p3) node[midway, above, font=\scriptsize] {propagates};

  \draw[black!20, thin] (-1.0,0.4) -- (8.8,0.4);

  \node[font=\small\bfseries, green!45!black] at (5.0,-0.25) {With reactive trigger};
  \node[box={orange!18}] (q1) at (0,-1.2)    {paper change\\+ fabrication};
  \node[box={green!18}]  (q2) at (5.0,-1.2) {Ground\\(forced)};
  \node[box={green!18}]  (q3) at (10.0,-1.2) {Audit\\(forced)};
  \draw[good] (q1) -- (q2);
  \draw[good] (q2) -- (q3);
  \node[font=\small, green!40!black, align=center] at (10.0,-2.25)
    {fabrication caught\\before propagation};
\end{tikzpicture}
\caption{Reactive grounding trigger. Any change to the paper or README forces an immediate grounding-then-audit pass, preventing hallucination accumulation from propagating beyond one step.}
\label{fig:trigger}
\end{figure}

After every step, the controller compares pre- and post-step git state.
If any file in $\mathcal{T} = \{\texttt{paper.tex},$ $\texttt{README.md}\}$ changed, it forces $(\pGroundingCreation, \pSkepticalAudit)$---grounding creation followed by skeptical audit (Figure~\ref{fig:trigger}).
The trigger exists because paper and README changes are the primary vector through which hallucinations enter the public projection~$P$: the LM writes a claim, the claim is not immediately checked, and subsequent prompts build on the fabricated claim as if it were true.
By forcing an audit within one step of any public-facing change, the trigger bounds the propagation depth of any single hallucination to at most one step.

\begin{proposition}[Bounded hallucination propagation]\label{prop:trigger}
Under the reactive grounding trigger, no public-facing artifact change can propagate through more than one step without an intervening grounding pass.
\end{proposition}

\subsection{Adjacency constraints}

The controller keeps the model from drifting too far in a single step. When an expansion prompt fires, the proposed pivot has to satisfy five small rules: it has to keep the project's main existing capability working; it has to be reachable in a single conceptual step from what the project already does, rather than a leap into an unrelated area; it has to be specific enough that the paper can describe one concrete instance of it; it has to make at least one of the existing benchmarks, tests, or grounding records stronger; and the new claim has to be backed by something the project can actually point to.
Every expansion is followed by the same grounding-then-audit pair as in the previous subsection.
Without these rules, the model tends to make large unjustified leaps---adding an entirely new feature domain that has no connection to the existing theory or benchmarks. Such leaps produce code with no benchmark to evaluate it on and paper claims with no theory behind them. Forcing each expansion to stay one step away from an existing capability, and to be checkable against existing evidence, keeps the repository growing in directions the rest of the project can actually support.

\section{Co-Evolution State Machine}\label{sec:formalism}

We formalize our controller as a \emph{co-evolution state machine}:
finite control modes encode the coarse development phase,
while a small auxiliary memory records unfinished quality debt
that should influence later prompt choice.
This choice is deliberate.
A plain finite-state controller is too weak for long-horizon repository work because it cannot remember that an expansive generation step created follow-up obligations for grounding, benchmarking, or paper repair.
The resulting formalism mirrors the code path in the implementation codebase:
a discrete prompt is selected, the corresponding prompt text
is executed by one or more agents against the repository,
and the workspace is then re-summarized from disk to obtain
the next observable state.

\begin{definition}[Co-Evolution State Machine]\label{def:co-evolution-sm}
A co-evolution state machine is a tuple $\AM = \langle q_0, Q,\; \Sigma,\; \tau : Q \to Q \rangle$ where
%
$Q = \W \times M \times \Reals_{\ge 0}^5 \times \Sigma^* \times \Sigma^*$ is the \emph{state space}.
A state $q_t = (W_t,\, m_t,\, \ObVec_t,\, F_t,\, H_t)$ at step $t$ consists of:
    \begin{itemize}[leftmargin=1.2em,itemsep=1pt]
      \item $W_t \in \W$: the workspace summary (\S\ref{sec:process}), capturing the current theory, repository forest, public projection (paper/README), evidence surface (benchmarks, tests, grounding ledger), utility hypothesis, and open obligations,
      \item $m_t \in M = \{\mathsf{S},\mathsf{G},\mathsf{H},\mathsf{T}\}$: the control mode (seed, generate, harden, tail),
      \item $\ObVec_t \in \Reals_{\ge 0}^5$: the obligation vector,
      \item $F_t \in \Sigma^*$: the forced follow-up queue,
      \item $H_t \in \Sigma^*$: the recent-prompt history.
    \end{itemize}
    $\Sigma$ is the 17-symbol prompt alphabet (Table~\ref{tab:prompt-families}).
    
    $\tau : Q \to Q$ is the \emph{transition map}, defined by the component maps $\scorer$, $\kappa$, $\delta$, $\alpha$ as follows.
    Given $q_t = (W_t, m_t, \ObVec_t, F_t, H_t)$ we compute the next state based on the operations:
    \papercomment{
    \begin{enumerate}[leftmargin=1.5em,itemsep=1pt]
      \item \textbf{Score.} The scorer $\scorer(W_t, \ObVec_t) \in \Reals^{|\Sigma|}$ assigns a real-valued score to each prompt symbol.
      \item \textbf{Filter.} The selection kernel $\kappa(m_t, W_t, H_t, \ObVec_t, F_t) \subseteq \Sigma$ returns the set of admissible prompts via mode guards, tail-budget overrides, and forced shells.
      \item \textbf{Select.} The executed prompt is $p_t \in \arg\max_{p \,\in\, \kappa(\cdots)} \scorer(W_t, \ObVec_t)_p$.
      \item \textbf{Update.} $(W_{t+1},\, m_{t+1}) = \delta(W_t, m_t, p_t)$, where $\delta : \W \times M \times \Sigma \to \W \times M$ re-reads the workspace from disk after the agent executes $p_t$ and advances the mode when phase guards are met.
      \item \textbf{Obligations.} $\ObVec_{t+1} = \decay \cdot \ObVec_t + \alpha(p_t)$, where $\alpha : \Sigma \to \Reals_{\ge 0}^5$ is the obligation-push map and $\decay = 2^{-1/8}$.
      \item \textbf{Side state.} $F_{t+1}$ and $H_{t+1}$ are recomputed from post-execution events (paper/README diffs, critique triggers, obligation-pop outcomes).
    \end{enumerate}
    }
    \[
    \begin{array}{rcll}
      P_t & = & \kappa(m_t, W_t, H_t, \ObVec_t, F_t) & \mbox{compute a set of admissible prompts} \\
      p_t & = & \arg\max_{p \,\in\, P_t} \scorer(W_t, \ObVec_t)_p & \mbox{selects prompt to be executed} \\
      (W_{t+1},\, m_{t+1}) & = & \delta(W_t, m_t, p_t) & \mbox{state after invoking agent iteration} \\
      \ObVec_{t+1} & = & \decay \cdot \ObVec_t + \alpha(p_t) & \mbox{decay based update to obligation} \\
      H_{t+1} & = & \mathrm{truncate}(H_t \cdot p_t) & \mbox{append $p_t$ and truncate to the fixed window} \\
      F_{t+1} & = & \mathit{FollowUpRule}[p_t] & \mbox{lookup into a static dictionary}      
    \end{array}
    \]
    where scorer $\scorer(W_t, \ObVec_t) \in \Reals^{|\Sigma|}$ assigns a real-valued score to each prompt symbol
    and the selection kernel $\kappa(m_t, W_t, H_t, \ObVec_t, F_t) \subseteq \Sigma$ returns the set of admissible prompts via mode guards, tail-budget overrides, and forced shells.
The initial state is $q_0 = (W_0, \mathsf{S}, \mathbf{0}, \varepsilon, \varepsilon)$.
The modes $M$ are the finite backbone; the workspace $W_t$ carries the evolving research-software state that the scorer and kernel condition on.
\end{definition}

The implementation factorizes controller evolution into a selection kernel and a post-execution update.
Given current mode $m_t$, workspace summary $W_t$, recent-prompt history $H_t$, obligation state $\ObVec_t$, and any forced follow-up queue $F_t$, the kernel $\kappa$ computes an admissible prompt set $P_t$ from structural guards, tail-budget overrides, and forced shells, then uses the linear scorer to select $p_t \in P_t$.
Crucially, selecting $p_t$ is not merely a symbolic transition: the controller dispatches the exact prompt associated with $p_t$ to an agent run, and that run edits files in the workspace.
The workspace evolves precisely through this execution step.
After the agent finishes, the update map $\delta$ forms $W_{t+1}$ by resummarizing the on-disk repository, appends $p_t$ to the truncated recent-prompt history, and recomputes controller side state such as the forced queue from post-step events including paper/README diffs, critique triggers, and obligation-pop outcomes.
So $H_{t+1}$ is not updated independently of the workspace: in the code it is a function of both the prior history and the observed post-execution repository state.

The decay factor $\decay = 2^{-1/8}$ gives a half-life of 8 steps, so
pressure $\pressure_t = \|\ObVec_t\|_1$ biases the scorer toward grounding when debts are high.
The exponential decay is also a design choice: fresh debts should matter strongly, but stale debts should not permanently dominate the controller once later steps have changed the project trajectory. This mirrors the role of discount factors in non-stationary multi-armed bandits, where exponential or sliding-window discounting keeps action values responsive to a drifting environment~\citep{garivier2011discounted,besbes2014nonstationary}.

\begin{proposition}[Bounded pressure]\label{prop:bounded}
$\pressure_t \le k \cdot \alpha_{\max} / (1 - \decay)$ for all $t$.
\end{proposition}

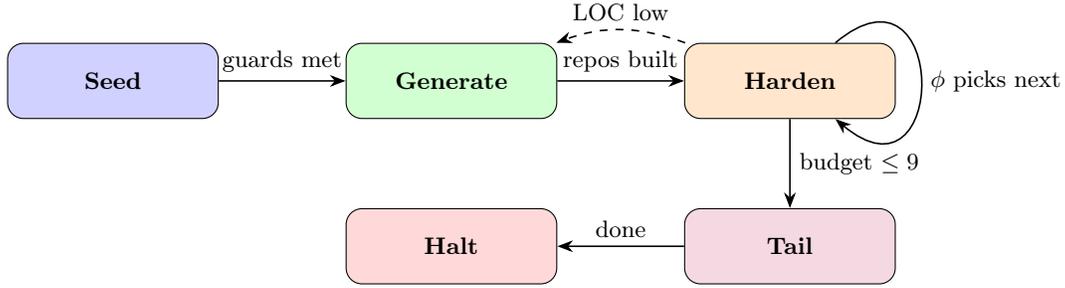
\begin{figure*}[t]
\centering
\begin{tikzpicture}[
  state/.style={draw, rounded corners=6pt, fill=#1,
                minimum width=2.8cm, minimum height=1.0cm,
                align=center, inner sep=8pt, font=\small\bfseries},
  trans/.style={-{Stealth[length=6pt]}, semithick},
]
  \node[state={blue!18}]   (seed) at (0,0)     {Seed};
  \node[state={green!18}]  (gen)  at (4.5,0)   {Generate};
  \node[state={orange!20}] (hard) at (9.0,0)   {Harden};
  \node[state={purple!15}] (tail) at (9.0,-2.2) {Tail};
  \node[state={red!15}]    (halt) at (4.5,-2.2) {Halt};

  \draw[trans] (seed) -- (gen)  node[midway, above, font=\small] {guards met};
  \draw[trans] (gen)  -- (hard) node[midway, above, font=\small] {repos built};
  \draw[trans] (hard) -- (tail) node[midway, right, font=\small] {budget $\le 9$};
  \draw[trans] (tail) -- (halt) node[midway, above, font=\small] {done};

  \draw[trans] (hard) to[out=40,in=-40,looseness=5]
    node[right, font=\small] {$\scorer$ picks next} (hard);

  \draw[trans, dashed, bend right=20] (hard) to
    node[above, font=\small] {LOC low} (gen);
\end{tikzpicture}
\caption{Controller mode graph (finite-mode backbone).
  Side state---obligation vector $\ObVec_t$, forced follow-up queue, and recent-prompt history---is not shown; see Figure~\ref{fig:architecture}.}
\label{fig:mode-graph}
\end{figure*}

This means no garbage collection, stack limits, or overflow handlers are needed.
The finite-mode backbone of this hierarchy is shown in Figure~\ref{fig:mode-graph}.

\paragraph{Prompt scoring.}
Each prompt symbol $p \in \Sigma$ has weight vector $w_p$ and bias $b_p$; the score is $\langle w_p, \mathrm{feat}(W_t, \ObVec_t)\rangle + b_p$.
Features include structural deficits, obligation pressure, LOC scaling pressure, and recency penalties.
This is deliberately a linear scorer---interpretability matters more than expressiveness in an auditable controller, because the point of the formalism is not merely to choose a prompt, but to make it legible why that prompt, and therefore why that exact agent action, was taken next.

\paragraph{A bandit view of the controller.}
It is useful to read the selection step as a small contextual bandit~\citep{robbins1952,auer2002finite,lattimore2020bandit}.
The 17 prompt families play the role of arms, the feature vector $\mathrm{feat}(W_t,\ObVec_t)$ plays the role of the context, and the linear score is the kind of action-value that linear contextual bandits use~\citep{li2010linucb,chu2011linucb}.
We never \emph{learn} this scorer; the weights are fixed by hand. The bandit framing is invoked only to read off two design choices.

First, the underlying ``rewards'' shift over the course of a run, because the prompts themselves rewrite the workspace they are scored against. This is the same problem that discounted or sliding-window UCB are designed for in non-stationary environments~\citep{garivier2011discounted,besbes2014nonstationary}. Our exponential decay $\decay = 2^{-1/8}$ on $\ObVec_t$ plays the same role: an unfinished follow-up loses half its weight every 8 steps, so the context the controller looks at reflects the current state of the project rather than its full history.

Second, we do not need a stochastic exploration term. Two earlier mechanisms already supply enough variety in what is admissible: the kernel restricts which prompts are even available, and the obligation vector itself raises the effective score of grounding and audit prompts whenever follow-ups have piled up (similar in spirit to how UCB inflates the value of under-pulled arms~\citep{auer2002finite}). The net controller stays within the bounded-state, auditable behaviour of a Mealy-style automaton while making each step's choice in a way a contextual bandit would recognise.

We view the combination---a hand-set linear scorer over workspace features, paired with a fading record of unfinished follow-ups---as a key contribution of \cometh{}: it is the smallest device we found that selects the next prompt from observable workspace state, handles long-horizon shifts without unbounded memory, and keeps every choice traceable to a short, readable feature vector.

\section{\athree{}: A flagship case study}\label{sec:a3}

We arrived at our automated methodology while developing \athree.
It represents the best-vetted repository so far.
\athree{} is a Python static-analysis tool available as a python wheel~\url{pypi.org/project/a3-python}.
For each warning it would otherwise raise, it tries to construct a small mathematical certificate---originally a method from control theory for proving that an unsafe state of a physical system is unreachable~\cite{prajna2004barrier}---and only keeps the warning if no such certificate can be found.
It augments this with two standard program-analysis ingredients: looking across function boundaries (interprocedural analysis~\cite{reps1995interprocedural}) and running the program symbolically along selected paths (dynamic symbolic execution~\cite{godefroid2005dart}).
The resulting tool combines several previously separate verification techniques in one pipeline, which is what makes its precision gain over each individual technique possible.
We discuss \athree{} in more depth online~\url{https://risemsr.github.io/blog/2026-02-16-halleyyoung-a3/}.
The trajectory of how it was developed illustrates both failure modes and how the controller addresses them.

\subsection{Theory evolution}

\athree{} passed through six revisions of its underlying theory (Table~\ref{tab:a3-epochs}). The first move (0$\to$1) replaced an early formalism that turned out to be too expensive to compute with safety certificates; without a controller mechanism that allows the formal goal to be revised in light of what is feasible, the model would have kept generating code under the original, unworkable specification. The later move (3$\to$4) followed a different pressure: by then the project had accumulated several previously independent verification techniques, and they had to be reorganised so that they cooperated rather than overlapped.

Both moves left the same 90-case benchmark in place throughout, so neither was free to drift arbitrarily.

\begin{table}[t]
\centering
\small
\caption{\athree{} loop epochs.}
\label{tab:a3-epochs}
\begin{tabularx}{\linewidth}{@{}>{\raggedright\arraybackslash}p{0.5cm}>{\raggedright\arraybackslash}p{3.3cm}X@{}}
\toprule
$k$ & \textbf{Discovery} & \textbf{Transition} \\
\midrule
0 & Quantitative safety & Distance-to-boundary framing \\
1 & Safety certificates & Prove unsafe states unreachable from reachable ones \\
2 & Executable Python & Bytecode control-flow graphs with exception edges \\
3 & Borrowed verification techniques & IC3/PDR~\cite{bradley2011ic3}, CHC/Spacer~\cite{komuravelli2016spacer}, CEGAR~\cite{clarke2000cegar}, ICE~\cite{garg2014ice} \\
4 & Combined pipeline & Shared producers of predicates, interpolants, certificates \\
5 & Evaluation survival & Only claims supported by benchmarks retained \\
\bottomrule
\end{tabularx}
\end{table}

\subsection{Evaluation}

\begin{table}[t]
\centering
\small
\caption{\athree{} evaluation results.}
\label{tab:a3-eval}
\begin{tabularx}{\linewidth}{@{}>{\raggedright\arraybackslash}p{3.4cm}>{\raggedright\arraybackslash}p{1.6cm}X@{}}
\toprule
\textbf{Surface} & \textbf{Metric} & \textbf{Result} \\
\midrule
90-case benchmark & P / R / $F_1$ & $.704 / .844 / .768$; next-best $.364$ \\
$-$Combined pipeline & $\Delta F_1$ & $-.052$, 6 extra FPs \\
$-$Interprocedural & $\Delta F_1$ & $-.012$, 3 missed TPs \\
$-$DSE & $\Delta F_1$ & $-.009$, unresolved recursion \\
\texttt{requests} scan & reduction & 183 $\to$ 4 confirmed TPs \\
LLM2CLIP scan & cascade & 55 $\to$ 5 confirmed TPs \\
Adafactor & safety & all 21 proven safe \\
\bottomrule
\end{tabularx}
\end{table}

On a 90 case-benchmark, several findings emerged:
\begin{itemize}
    \item The ablations summarized in Table~\ref{tab:a3-eval} are \emph{monotone}: each added layer helps, none hurts.
    \item We take this as evidence that the controller's theory-revision mechanism produces load-bearing internal structure, not decorative additions.  The large increase in $F_1$ over any extant tool indicates that the theory-code coevolution truly lead to a breakthrough in the problem of python bugfinding, a highly challenging task and one that has resisted prior automated approaches.  
\end{itemize}

Furthermore, \athree{} proved useful in practice.  When pursuing bugfinding on candidate Python packages, it was able to reduce the number of potential true positives from hundreds to a handful of confirmed issues, which, when given to an LLM for further analysis, were often (but not always) correctly identified as genuine bugs.  The improvement on False Positives was largely a product of DSE, a late stage evolution in the theory and code.  LLM2Clip had one particularly noteworthy found and confirmed bug (a bug which had already been fixed in the official PyTorch implementation of Adafactor, but was not yet implemented for this repo).  This demonstrates that \athree{} can effectively prioritize and filter potential issues, making it a valuable tool for practical software verification tasks.
\section{Portfolio and evolution}\label{sec:evaluation}

The pipeline has produced 46 repositories across 12+ domains (Appendix~\ref{app:portfolio}).

The orchestrator does three things consistently across the portfolio. When one of its intended claims turns out not to hold, it retracts that claim and looks for a different angle from which the project can still be state-of-the-art, rather than quietly reporting the original claim anyway. When the budget still has steps left and the project has already done what it originally set out to do, it broadens its scope to a nearby, larger problem instead of stopping. And when neither of those is enough, it revises not only the code but the underlying theory itself, so the artifact and the story it tells stay in sync.

In \athree, the starting idea is small and practical: most warnings a Python static analyzer raises do not actually correspond to bugs, so a developer ends up wading through false alarms. The project asks whether the loop can do that triage on its own. Over the run, the system shifts from generic ``check the code for problems'' toward something more specific: it learns to build a little safety argument for each warning, asks an SMT solver to confirm that argument, and only flags what survives. By the end the repository is no longer presented as ``a Python static analyzer'' in the abstract; it is presented, and benchmarked, as a precision-focused bug filter, with $F_1{=}0.768$ against a $0.364$ baseline.

In \texttt{dp-mech-forge}, the starting idea is sharper to begin with: instead of designing differential-privacy mechanisms by hand, treat the design problem as a search loop where a candidate mechanism is proposed and then a verifier either accepts it or returns a concrete counterexample to learn from. Generation grows that loop outward---into approximate-DP, streaming data, workload-aware optimization, several solver backends, and accounting that respects composition---without ever abandoning the propose/verify shape. Hardening adds machine-checkable certificates, numerical safety nets, and a benchmark ledger that reports many configurations rather than a single hero result. The surviving claim is a measured accuracy improvement, not a sweeping ``we redesigned DP'' story: a median $3.66\times$ utility gain over Laplace, reaching roughly $80\mathrm{K}\times$ in the high-privacy regime at $\varepsilon{=}0.01$.

In \texttt{cross-lang-verifier}, the starting observation is that automatically translated C-to-Rust code can pass every test you run against it and still be wrong, because the bug lives in what the C standard leaves undefined rather than in any visible output. The loop turns that observation into a pipeline that compares the two programs against a shared semantic reference, hunts for inputs on which they would diverge, and---when it finds one---feeds that example back to repair the translation rather than just rejecting it. The README, CLI traces, and benchmarks are then rewritten so the project promises only what it actually demonstrates: full success on the targeted undefined-behavior benchmark slice and $F_1{=}0.968$ after the public claims are tightened.

In \texttt{tensorguard}, the starting idea is the everyday frustration that a PyTorch model crashes deep inside training because two tensors disagree about their shape. The natural goal is to catch this without running the code. Generation pushes well past plain shape checking: the same checker also tracks which device a tensor lives on, whether the model is in train or eval mode, the memory layout, and how dimensions get permuted, and it learns new rules when it gets caught out by a counterexample. Hardening narrows the public framing to what the artifact actually supports---a strict mode that is conservative but reliable, a more permissive mode with moderate recall, integration with standard developer tooling, and a Lean~\cite{moura2021lean4} formalization of the core rules---rather than an over-broad ``universal tensor verifier'' claim.

In \texttt{ml-pipeline-selfheal}, the starting goal is more ambitious than the final system keeps: catch and fix ML-pipeline bugs purely from the source. The loop is forced to back off and let the running pipeline itself participate---watching for anomalies, diagnosing what likely caused them, proposing a fix, replaying the suspect computation against that fix, and escalating to a human when none of that is safe. The hardening phase here is unusually audit-heavy: out-of-distribution stress tests, replay checks, TLA+ exploration of the controller's states, and explicit publication of the cases the system refuses to handle on its own. What survives is a deliberately bounded self-healing claim, not an autonomy claim: $85.7\%$ repair on the controlled benchmark, with a frankly reported $9.1\%$ failure rate out of distribution.

Taken together, these trajectories show the same portfolio regularity. The first idea is not thrown away; it is grown outward into a richer system, and then trimmed back by benchmarks, audits, and honestly reported failures until what remains is smaller, sharper, and harder to fake. We see similarities in stages across all repositories in the portfolio: a generation phase that drifts to neighboring problems, followed by a contraction phase that decides which of those drifts are allowed to stand.

\section{Observations about LM behavior}\label{sec:observations}

The roughly 400 commits in the portfolio give a concrete view of how the LM behaves during a long, controller-driven development process.
We make four observations from that history.
These are descriptive patterns we saw in the surviving git record, not causal claims.
Table~\ref{tab:observations} summarizes the main evidence and takeaway for each one.

\begin{table}[t]
\centering
\small
\caption{Summary of observations on LM behavior across the 46-repository portfolio.}
\label{tab:observations}
\begin{tabularx}{\linewidth}{@{}>{\raggedright\arraybackslash}p{0.4cm}>{\raggedright\arraybackslash}p{2.3cm}>{\raggedright\arraybackslash}p{2.8cm}>{\raggedright\arraybackslash}X@{}}
\toprule
 & \textbf{Observation} & \textbf{Evidence} & \textbf{Implication} \\
\midrule
1 & Audit dominates late phases & Majority of late commits are grounding/repair, not features & Forced grounding shell does real work \\
2 & Theory revision is possible when forced & \athree{} made two major pivots under adjacency support & LMs default to desynchronization without explicit controller support \\
3 & Visible failure beats concealed success & \athree, dp-mech-forge, ml-pipe-selfheal publish honest gaps & Audit pass surfaces limitations the LM would otherwise hide \\
4 & LMs self-organize toward compositionality & Submodule conversion, kitchensink epoch & Emerges from development horizon, not from prompting \\
\bottomrule
\end{tabularx}
\end{table}

\paragraph{Observation~1: Audit dominates late trajectories.}
Across the portfolio, most late-stage commits are about checking claims, grounding results, and fixing mistakes---not adding new features.
Commit messages such as \textit{``Fix hallucination issues''} and \textit{``Add honest empirical status''} line up directly with the grounding and audit prompts ($\pGroundingCreation$, $\pSkepticalAudit$).
The forced grounding shell is doing real work: without it, the LM tends to keep producing new features instead of stopping to verify what it has already said.

\paragraph{Observation~2: LMs are good at theory revision when forced.}
\athree's two major pivots show that an LM can substantially revise a mathematical framework when the controller explicitly supports that move.
Without that support, the model usually keeps optimizing for the original plan, even when that plan no longer fits the evidence.
The adjacency mechanism helps because it treats revision as an allowed, rewarded action instead of as a mistake.

\paragraph{Observation~3: Visible degradation beats concealed success.}
The most trustworthy repositories are the ones that openly state what did not work or what remains uncertain.
For example, \athree{} keeps \texttt{UNKNOWN} verdicts for recursive cases, \texttt{dp-mechanism-forge} reports a Laplace fallback when synthesis paths are unavailable, and \texttt{ml-pipeline-selfheal} reports OOD detection at 9.1\%.
This behavior was not hand-designed into the controller.
It emerged from the skeptical audit pass ($\pSkepticalAudit$), which pushes the LM to mark places where its claims go beyond the available evidence.

\paragraph{Observation~4: The model self-organizes toward compositionality.}
In the middle of the development trajectories, several commits reorganize the portfolio into cleaner modules, including \textit{``Convert subdirectories to git submodules''} and \textit{``Rename all repos to short descriptive names''}.
In \athree, the kitchensink epoch similarly reorganized imported techniques as typed producers.
Taken together, these cases suggest that LMs can discover more compositional structure over a long enough development horizon, even though they usually do not move in that direction on their own.

\section{Controller-level ablations}\label{sec:ablations}

The formalization induces six ablations that operate on the orchestration law.
They are shown in Table~\ref{tab:ablations}.

\begin{table}[t]
\centering
\small
\caption{Controller-level ablations. Each removes a structural component and predicts a specific degradation.}
\label{tab:ablations}
\begin{tabularx}{\linewidth}{@{}>{\raggedright\arraybackslash}p{4.0cm}>{\raggedright\arraybackslash}p{4.6cm}X@{}}
\toprule
\textbf{Ablated} & \textbf{What changes} & \textbf{Predicted signature} \\
\midrule
Adjacency & theory frozen after seed & fewer pivots, weaker benchmark gains \\
Grounding ledger & no audited-claim tracking & larger unsupported claim surface \\
Paper-first & no early paper skeleton & later synchronization of $P$ and $E$ \\
Obligation decay & debts never fade & unbounded pressure, brittleness \\
Grounding trigger & no reactive audit & hallucinations persist across steps \\
Bench.\ judge & no oracle search & weaker evaluation surfaces \\
\bottomrule
\end{tabularx}
\end{table}

These are \emph{predicted} by the formalism, not post-hoc.
The first three test process design; the last three test controller mechanism.
Full experimental validation across batch runs is ongoing work.

\section{Discussion and limitations}\label{sec:discussion}

\paragraph{Research as co-evolution.}
Treating research as a co-evolution problem changes what counts as a good controller.
A controller should not be judged only by whether it can produce code quickly or write plausible papers.
It should be judged by whether it keeps the coupled artifact state in a region where claims remain supportable, benchmarks remain representative of the current implementation, and theory continues to describe what the software actually does.
This is why a purely task-completion view is too weak for research software: local success can increase global incoherence.
An LM can improve a benchmark section, refactor code, or sharpen a theorem statement in ways that look productive in isolation while making the overall project less scientifically trustworthy.

This perspective also changes the meaning of iteration.
In ordinary software automation, iteration is often modeled as repeated improvement of one artifact against a fixed objective.
In research, the objective itself is partially endogenous: the problem statement, the mathematical framing, the benchmark protocol, and even the headline claim can all be revised in response to what the implementation reveals.
The important consequence is that \emph{theory is a mutable coordinate}, not merely an upstream input.
LMs are increasingly used to formulate problems, invent abstractions, and rewrite explanations, so a controller that assumes the theory surface is fixed will systematically miss the moments where the right action is to revise the research question rather than continue optimizing stale code.

One ramification is methodological.
If research is co-evolutionary, then evaluation should focus less on end-state artifact quality alone and more on state-management properties throughout the run.
Questions such as ``Did the controller ground new claims immediately after expansion?'', ``Did benchmark pressure fire when evidence lagged?'', and ``Did theory revision follow implementation discoveries rather than precede them dogmatically?'' become first-class evaluation criteria.
This suggests that future work on LM research agents should measure stability, recoverability after drift, and the latency between artifact mutation and integrity repair, not just final benchmark scores or paper quality.

Another ramification is organizational.
Once research is viewed as co-evolution, familiar boundaries between ideation, implementation, evaluation, and writing become less clean.
Those activities are not separate pipeline stages with complete handoffs; they are coupled maintenance operations on a shared evolving state.
Recent systems such as Ralph-style engineering loops, AutoResearchClaw-style staged research pipelines, and Deep Research-style asynchronous investigation suggest that progress is increasingly being made at the level of process architecture rather than isolated prompting tricks or single-step task performance.
That makes explicit controller structure more important, not less.
Without typed obligations, guarded transitions, and forced repair passes, an LM system tends to overproduce attractive surface artifacts while underinvesting in the repair work that keeps those artifacts mutually consistent.
In that sense, obligation tracking and reactive grounding are not just practical heuristics; they are a minimal governance layer for research automation.

\paragraph{Limitations.}
We mitigate confirmation bias through externally runnable benchmarks, machine-readable grounding ledgers, and falsifiable ablation predictions.
\athree{} is the most extensively tested case; the remaining 45 repositories vary in how much analysis by domain experts has been conducted.
More broadly, the we argue for a structural view of research automation, though the evidence for
optimality or universality of controller patterns in the co-evolution state machine
is at best based on what we could check empirically.
Different research domains may require different couplings, different integrity triggers,
or stronger forms of symbolic state than the co-evolution state machine.

\section{Related work}\label{sec:related}

\paragraph{Program-search and optimization.}
FunSearch~\citep{romera2024funsearch} and AlphaEvolve~\citep{novikov2025alphaevolve} optimize programs inside a fixed evaluation frame.
The Darwin G\"odel Machine~\citep{zhang2025dgm} extends this to self-modifying code but still assumes a fixed utility metric.
\cometh{} differs in that the evaluation frame itself is under construction.

\paragraph{Repository-scale agents.}
SWE-agent~\citep{yang2024sweagent}, Devin~\citep{cognition2024devin}, ChatDev~\citep{qian2024chatdev}, and MetaGPT~\citep{hong2024metagpt} operate on shorter horizons and evaluate success as patch correctness or prototype completion.
Ralph-style persistent engineering loops~\citep{huntley2026ralph} extend this family from single tasks to continuous repository-local iteration with repeated validation and spec-backed memory.
\cometh{} inherits that long-horizon repository loop, but treats the thesis, benchmark suite, and grounding surface as mutable coordinates of state rather than as a backlog to be driven to closure.

\paragraph{Automated research.}
The AI Scientist~\citep{lu2024aiscientist} and SciAgents~\citep{SciAgents} automate ideation, experiment planning, and paper drafting but operate closer to idea space than code space.
AutoResearchClaw~\citep{liu2026autoresearchclaw} is a closer recent neighbor: it combines literature search, experiment design, execution, paper drafting, verification, and human intervention modes in a single staged pipeline.
OpenAI Deep Research~\citep{openai2025deepresearch} contributes another adjacent capability: asynchronous multi-step online investigation with citation-backed synthesis.
\cometh{} extends this line by making the primary object of control not a paper-production pipeline or a research report, but the coupled evolution of theory, code, evidence, and claims.
Its unit of output is a repository whose code, README, benchmark harnesses, and mathematical story must repeatedly re-align under evidence.

\paragraph{Controller formalisms.}
Most agent orchestration uses flat FSMs, ad-hoc heuristics, or learned policies.
ReAct~\citep{yao2023react} interleaves reasoning and action but has no persistent obligation tracking.
Reflexion~\citep{shinn2023reflexion} adds self-reflection but no formal model of debt accumulation and decay.
Voyager~\citep{wang2023voyager} maintains a skill library but no workspace-state model.
Our augmented Mealy machine with decaying obligation memory offers a principled middle ground: enough memory to track quality debts, enough flexibility to tolerate stochastic failures, and provably bounded internal state.

\paragraph{Multi-armed bandits and prompt selection.}
The selection step in our controller---picking $p_t \in \arg\max_{p \in P_t}\scorer(W_t,\ObVec_t)_p$---is a non-stationary contextual multi-armed bandit. Multi-armed bandits date back to~\citet{robbins1952} and were given finite-time analysis by~\citet{auer2002finite}; \citet{lattimore2020bandit} survey the modern theory. Contextual bandits with linear value functions, in particular LinUCB and its disjoint variant~\citep{li2010linucb,chu2011linucb}, are the closest classical analogue of our linear scorer over workspace features. Non-stationary settings, where arm rewards drift over time, are typically handled by discounted-reward UCB or sliding-window UCB~\citep{garivier2011discounted,besbes2014nonstationary}; our exponentially decaying obligation vector $\ObVec_t$ is the corresponding device for long-horizon LM orchestration. A growing line of work uses bandit algorithms inside LM systems---for example, casting prompt or example selection as a best-arm-identification problem~\citep{shi2024prompt}---typically optimising a single scalar reward such as accuracy. \cometh{} differs in three respects: (i) the ``reward'' is a vector of artifact deficits rather than a scalar score, (ii) non-stationarity is not a nuisance but the dominant signal, encoded explicitly through the decaying obligation vector, and (iii) we deliberately keep the policy a fixed, hand-specified linear scorer rather than a learned one, because in research-software construction interpretability and per-step auditability are more important than asymptotic regret.

\section{Conclusion}

Research software development occupies an under-served regime where mathematical theory, code, benchmarks, and public claims must mature together under an evolving research question and/or agenda.
We identified two LM-specific failure modes---hallucination accumulation and desynchronization---and proposed \cometh, an iterative prompt automaton that addresses them through coupled artifact surfaces, decaying obligation memory, and reactive grounding triggers.

The pipeline has produced 46 repositories across 12+ domains.
The flagship \athree{} reaches $F_1 = 0.768$ vs.\ $0.364$ with monotone ablation gains.
Analysis of ${\sim}400$ commits reveals that audit and claim contraction dominate late trajectories, hallucination repair is a persistent multi-layer challenge, and LMs can discover compositional structure when given sufficient development horizon.
The co-evolution problem is a fundamental challenge for LM-based research; the mechanisms introduced here are a first step toward principled solutions.

\newpage
\bibliographystyle{plainnat}
\bibliography{refs}


\clearpage
\appendix

\section*{Note on LLM Usage}

Portions of the writing in this paper were drafted with the assistance
of large language models (specifically, drafts were generated using
GitHub Copilot and Claude). All generated text was carefully read,
edited, and vetted by the authors; no content was included without
human review. The research ideas, formal definitions, experimental
design, and conclusions are entirely the authors' own.

\section{Full portfolio listing}\label{app:portfolio}

The 46 repositories, grouped by domain.

\begingroup
\sloppy
\emergencystretch=3em
\hyphenpenalty=50
\exhyphenpenalty=50
\renewcommand{\ttdefault}{cmtt}

\begin{description}[leftmargin=1em, labelindent=0pt, font=\normalfont\bfseries, itemsep=8pt]

\item[Formal verification]\hfill
\begin{description}[leftmargin=1.5em, labelindent=0pt, font=\normalfont\ttfamily, itemsep=8pt]

  \item[txn-isolation-verifier]
  \normalfont
  Most production databases promise an ``isolation level'' (read committed, snapshot, serializable, etc.)~that bounds how transactions running in parallel may interfere. In practice the same nominal isolation level often behaves differently across SQL engines, and bugs at this layer cause silent data corruption that is extremely hard to reproduce. \texttt{txn-isolation-verifier} treats this as a formal-methods problem: it encodes each engine's claimed semantics as logical constraints, hands them to an SMT solver, and asks the solver to synthesise small transaction schedules---``anomaly witnesses''---on which two engines disagree. The output is a portable test suite of concrete, executable counterexamples that pinpoint exactly where one engine's behaviour drifts from another's.

  \item[cascade-config-verifier]
  \normalfont
  Real outages in microservice systems are often not single failures but \emph{cascades}: a slow downstream call triggers retries, retries amplify load, timeouts compound, and the whole system tips over. \texttt{cascade-config-verifier} tries to catch these failure modes from configuration files alone, before deployment. It builds a Retry/Timeout Interaction Graph (RTIG) from service configs, encodes it in TLA+ and Coq, and proves bounds on retry amplification and timeout-chain depth. For someone who normally thinks of configuration as ``just YAML'', the contribution is that the YAML can be checked for safety properties in the same way we check code.

  \item[tlaplus-coalgebra-compress]
  \normalfont
  Model checkers like TLA+ verify distributed algorithms by exhaustively exploring their reachable states; the bottleneck is almost always state-space size. \texttt{tlaplus-coalgebra-compress} uses a category-theoretic technique called F-coalgebra bisimulation to detect states that are observationally equivalent and quotient them away before search. The result is a drop-in pre-processor that shrinks TLA+ specs while preserving exactly the properties the user wanted to check. The takeaway for a non-specialist is that fairly abstract mathematics (coalgebra) translates here into a concrete speedup for an industrially used verification tool.

  \item[tensorguard]
  \normalfont
  Anyone who has trained a deep network has seen runtime crashes from tensor shapes that don't line up, tensors silently living on the wrong device, or operations called in the wrong training/eval phase. \texttt{tensorguard} is a static analyser for PyTorch that catches these errors before the model ever runs. It encodes shape, device, and training-phase information as Z3 refinement types and propagates them through the program, flagging contradictions as type errors. To an outside reader: it does for tensor programs roughly what a strict type checker does for ordinary code.

  \item[litmus-inf]
  \normalfont
  CPUs and GPUs do not actually execute memory operations in source order; each architecture defines a \emph{memory model} describing which reorderings are legal. ``Litmus tests'' are tiny multi-threaded programs designed to expose these reorderings. \texttt{litmus-inf} verifies large suites of such tests across architectures (x86, ARM, GPU) by exploiting the algebraic symmetries between threads to compress the search space. For a graphics or systems programmer, the value is a single tool that says, with proof, ``this concurrent code pattern is portable'' or ``this one is not''.

  \item[fp-diagnosis-repair-engine]
  \normalfont
  Floating-point arithmetic in scientific code looks innocent but accumulates rounding errors that can dominate the final answer. \texttt{fp-diagnosis-repair-engine} both \emph{diagnoses} which lines of a numerical pipeline contribute most to the error and \emph{repairs} them by synthesising a higher-precision implementation that meets a user-specified accuracy bound. Internally it combines interval arithmetic with program synthesis. The intuition for outsiders: it's the floating-point analogue of a profiler that not only locates the hot spot but also rewrites it for you.

  \item[mutation-contract-synth]
  \normalfont
  ``Mutation testing'' perturbs a program (e.g., flipping a $<$ to a $\le$) to check whether the test suite catches the change; surviving mutants suggest weak tests. \texttt{mutation-contract-synth} flips this on its head: it uses the pattern of which mutants survive to \emph{infer formal contracts} (pre/postconditions, invariants) that the original program must have implicitly relied on. The system rests on a duality between mutations and specifications. For a working developer, the practical effect is automatically generated, machine-checkable specs derived just from existing tests.

\end{description}

\item[Differential privacy]\hfill
\begin{description}[leftmargin=1.5em, labelindent=0pt, font=\normalfont\ttfamily, itemsep=8pt]

  \item[dp-mechanism-forge]
  \normalfont
  Differential privacy (DP) hides individuals in a dataset by adding carefully calibrated noise to released statistics. Choosing \emph{which} noise distribution to use, and how to scale it, is normally a manual derivation by an expert. \texttt{dp-mechanism-forge} treats mechanism design as an optimisation problem: given a query workload and a privacy budget, it searches for a noise mechanism that minimises expected utility loss and emits a machine-checkable certificate that the result is provably optimal in a formally specified sense. For a non-DP person, this is essentially ``compiler for privacy mechanisms''.

  \item[dp-verify-repair]
  \normalfont
  Even ``standard'' DP algorithms in the literature have repeatedly been found to have subtle bugs that violate the privacy guarantee. \texttt{dp-verify-repair} symbolically checks a candidate mechanism against six common DP definitions (pure, approximate, R\'enyi, zCDP, etc.), and when it finds a counterexample, it synthesises the smallest patch to the code that restores correctness. To an outside reader: it is a static analyser specialised to a property---privacy---that ordinary type systems cannot see.

  \item[certified-leakage-contracts]
  \normalfont
  When cryptographic code runs on a real CPU, the cache leaks information about which memory addresses were accessed, and an attacker can sometimes recover secret keys from this side channel. \texttt{certified-leakage-contracts} analyses x86-64 binaries directly and computes a tight upper bound on how many bits of secret information the cache pattern can reveal, together with a machine-checkable proof of that bound. The point for a non-security person: every binary gets a numeric ``leakage budget'' you can cite in a security argument rather than waving hands at constant-time discipline.

\end{description}

\item[Causal inference]\hfill
\begin{description}[leftmargin=1.5em, labelindent=0pt, font=\normalfont\ttfamily, itemsep=8pt]

  \item[causal-plasticity-atlas]
  \normalfont
  A causal model trained in one environment (a hospital, a market, a robot domain) often transfers poorly to another. \texttt{causal-plasticity-atlas} systematically maps which mechanisms in a causal graph are \emph{invariant} across environments and which are \emph{plastic}. It uses quality-diversity search (similar in spirit to evolutionary algorithms that maintain diverse populations) to fill an archive of environments and report, per edge of the graph, how robust that causal relationship is. The result is a kind of map of which parts of a causal model you can trust outside the data you trained on.

  \item[causal-qd-illumination]
  \normalfont
  In causal discovery, many different graphs are statistically indistinguishable from the data---they form a Markov equivalence class. Standard algorithms return one representative; \texttt{causal-qd-illumination} uses a quality-diversity algorithm called MAP-Elites to enumerate a diverse set of graphs that all explain the data, ``illuminating'' the equivalence class. For a downstream user, this turns causal discovery from a single point estimate into an explorable atlas of plausible causal stories.

  \item[causal-risk-bounds]
  \normalfont
  In financial networks, ``what if Bank A defaults?'' is a causal question that classical statistics underestimates because feedback loops are hard to identify from data alone. \texttt{causal-risk-bounds} encodes Balke--Pearl bounds (linear programs over causal models) and adversarial Monte Carlo Tree Search to compute Z3-verified upper and lower bounds on systemic risk. The output is not a point estimate but a certified interval, which a regulator can use as a worst-case guarantee.

  \item[causal-robustness-radii]
  \normalfont
  Causal DAGs are usually drawn as if their edges were known with certainty, but in practice each edge is estimated from noisy data. \texttt{causal-robustness-radii} computes, for each edge, a ``robustness radius'': how much you can perturb the data before that edge flips or disappears. The implementation is essentially sensitivity analysis, but packaged so that any DAG learned by another tool can be stress-tested. For someone outside causal inference, the takeaway is a fragility score per claim rather than a binary ``the edge is in the model''.

  \item[causal-trading-shields]
  \normalfont
  Quant strategies based on machine learning sometimes look great in backtests and then blow up live, partly because the learned model has no mechanism to refuse risky actions. \texttt{causal-trading-shields} couples a causal-discovery front end with a PAC-Bayes safety shield expressed as an interval Markov decision process; it formally verifies that the shielded strategy stays within preset risk envelopes regardless of which model in a posterior set is correct. The high-level idea: a ``please don't bankrupt us'' filter with a real proof attached.

\end{description}

\item[Program analysis]\hfill
\begin{description}[leftmargin=1.5em, labelindent=0pt, font=\normalfont\ttfamily, itemsep=8pt]

  \item[cross-lang-verifier]
  \normalfont
  Translating C code to Rust is a popular way to ``modernise'' legacy systems, but the translation can subtly change behaviour in ways that compiler-IR tools (like LLVM-based checkers) do not see, because they look at the lowered intermediate representation rather than at language-level semantics. To a graphics person, the analogy is checking that two shaders compute the same image even though their HLSL source uses very different control flow---looking only at the GPU machine code may miss the discrepancy. \texttt{cross-lang-verifier} works at the C and Rust source level: it builds matched abstract models of both programs, compares them under the languages' \emph{undefined-behaviour} rules, and reports cases where the C program is well-defined but its Rust translation is not (or vice versa). The deliverable is a list of concrete inputs on which the two programs are allowed to disagree, so engineers porting code can decide whether the divergence is a bug or an intended cleanup.

  \item[ml-pipeline-leakage-auditor]
  \normalfont
  ``Data leakage'' in machine learning is when information from the test set sneaks into the training pipeline (e.g., feature scaling fitted on the union of train and test), giving falsely optimistic accuracy. \texttt{ml-pipeline-leakage-auditor} is essentially a taint tracker for Python ML code: it labels test data as a tainted source, propagates the taint through pandas/scikit-learn operations, and flags every place where tainted bits influence training-time computation. For a non-ML reader, it is information-flow analysis applied to scientific Python.

  \item[ml-pipeline-selfheal]
  \normalfont
  Long-running ML pipelines fail \emph{silently}: data drifts, a feature stops being computed, accuracy slowly slumps, and no exception is ever raised. \texttt{ml-pipeline-selfheal} is a runtime that watches a pipeline's signals, diagnoses which stage is misbehaving, and performs a small repair (re-fit a normaliser, fall back to a simpler model, etc.) under a Lyapunov-style stability argument that proves the recovery process itself converges. The framing for outsiders: a control loop with a formal stability guarantee, wrapped around a data-science pipeline.

  \item[algebraic-repair-calculus]
  \normalfont
  When a single input file in a data pipeline changes, you usually do not want to re-run the whole pipeline. \texttt{algebraic-repair-calculus} provides an algebraic semantics with three sorts (data, deltas, repairs) that lets the system compute the minimal cost-optimal patch to apply downstream. It generalises traditional incremental view maintenance from databases to arbitrary user-defined operators. To a non-database reader, it is roughly ``\texttt{make}, but provably propagating the smallest correct change''.

\end{description}

\item[Mechanism design]\hfill
\begin{description}[leftmargin=1.5em, labelindent=0pt, font=\normalfont\ttfamily, itemsep=8pt]

  \item[algo-collusion-certifier]
  \normalfont
  Pricing algorithms used by online retailers can, without any explicit communication, learn to mimic cartel behaviour---raising prices in lockstep. Antitrust authorities currently struggle to prove this from black-box behaviour. \texttt{algo-collusion-certifier} probes such an algorithm with synthetic market histories, fits a behavioural model, and emits a machine-checkable certificate that the algorithm's policy is consistent (or inconsistent) with collusive equilibria. For someone outside economics, this is automated red-teaming with a formal artefact regulators can take to court.

  \item[market-manipulation-prover]
  \normalfont
  Compliance teams flag suspicious trading patterns daily, but turning an alert into a legal case requires a precise formal argument that the activity matches a manipulation pattern (spoofing, layering, etc.). \texttt{market-manipulation-prover} encodes manipulation patterns as SMT formulas, checks observed order books against them, and produces Z3/CVC5 proof certificates. The deliverable is an alert plus a formal proof object, so the alert is legally defensible rather than a black-box score.

  \item[bounded-rational-usability-oracle]
  \normalfont
  Usability regressions---a button moving, a workflow gaining a step---are usually caught only when users complain. \texttt{bounded-rational-usability-oracle} models a user as an information-theoretic bounded-rational agent (think: a noisy KL-regularised decision maker) and computes, for each candidate UI change, how much harder the modelled user has to work to complete tasks. Plugged into CI/CD, it acts like a unit test that fails when the UI gets noticeably less usable, without needing live A/B testing.

\end{description}

\item[ML systems]\hfill
\begin{description}[leftmargin=1.5em, labelindent=0pt, font=\normalfont\ttfamily, itemsep=8pt]

  \item[nn-init-phases]
  \normalfont
  Modern deep networks have many ways to fail at initialisation: dead ReLU neurons, exploding gradients, signal collapse, miscalibrated normalisation. Practitioners usually catch these by ``looking at the loss curve''. \texttt{nn-init-phases} formalises a taxonomy of 49 initialisation pathologies and tests each of them automatically across a model zoo, reporting a 96\% detection rate. For a non-ML reader, it is a static (and short-dynamic) checker for neural networks comparable to a linter for ordinary code.

  \item[diversity-decoding]
  \normalfont
  When you sample many outputs from an LLM (for re-ranking, ensembling, or brainstorming), you want them to be diverse, not near-duplicates. \texttt{diversity-decoding} formalises ``maximally diverse subset'' selection over a candidate pool, encodes it as an SMT problem, and returns a subset together with a certificate of optimality given the chosen diversity metric. The intuition for outsiders: it replaces ad-hoc heuristics like nucleus sampling plus deduplication with a small optimisation problem with a proof.

  \item[rag-fusion-compiler]
  \normalfont
  Retrieval-augmented generation (RAG) pipelines are a sequence of retrieval, re-ranking, filtering, and generation stages, each often re-querying the same store with overlapping queries. \texttt{rag-fusion-compiler} treats the pipeline as a relational query and applies operator-fusion-style rewrites that preserve recall but eliminate redundant retrievals, achieving roughly $5\times$ end-to-end speedup. To a database reader this is a query optimiser; to an ML reader it is a way to make production RAG cheaper without changing the model.

  \item[sparse-cpu-inference]
  \normalfont
  Mixture-of-Experts (MoE) language models activate only a few experts per token, but mainstream inference engines still pay full dense cost on CPU. \texttt{sparse-cpu-inference} is a speculative inference engine that predicts which experts will be needed, prefetches their weights, and skips the rest, achieving roughly $6.6\times$ speedup on representative MoE models. For a non-ML systems engineer, it is a clean example of branch prediction and speculative execution applied at the level of model structure rather than instructions.

\end{description}

\item[Distributed systems]\hfill
\begin{description}[leftmargin=1.5em, labelindent=0pt, font=\normalfont\ttfamily, itemsep=8pt]

  \item[safe-deploy-planner]
  \normalfont
  Staged rollouts (canary, blue/green) protect users by deploying new code to a fraction of traffic first, but ``how much can we shift, and when do we abort?'' is usually decided by intuition. \texttt{safe-deploy-planner} models the rollout as a control problem with measurable health signals and computes a \emph{rollback safety envelope}: the largest traffic share you may shift while still being able to revert within an SLA. The output is a deployment plan with formal pre- and post-conditions on each step, rather than a runbook the on-call has to interpret.

  \item[marl-race-detect]
  \normalfont
  Multi-agent reinforcement learning systems often look correct in unit tests but fail in production because of subtle interleavings between agents---an ordering of messages that the trainer never explored. \texttt{marl-race-detect} adapts classical race-condition analysis (lockset and happens-before) to MARL controllers, finding scheduler interleavings that violate stated safety invariants. For a non-RL systems person, it is essentially a Helgrind/TSan analogue for cooperating learning agents.

  \item[pram-compiler]
  \normalfont
  PRAM (parallel random-access machine) algorithms have beautiful work/span analyses on paper but rarely run on real hardware. \texttt{pram-compiler} compiles a library of 51 PRAM algorithms to executable hash-partitioned plans on commodity clusters and provides verified bounds on their work and span. To a non-theory reader, the contribution is a clean bridge between the parallel-algorithms textbook and code that actually runs.

  \item[proto-downgrade-synth]
  \normalfont
  Network protocols (TLS, SSH, application-level handshakes) are routinely attacked by tricking peers into negotiating an older, weaker version. \texttt{proto-downgrade-synth} reads a protocol library's source code, models its negotiation logic symbolically, and synthesises bounded sequences of messages that constitute a valid downgrade attack, if any exist. The output is concrete adversarial traces a developer can replay, rather than abstract advice to ``audit your handshake''.

\end{description}

\item[Bio/health]\hfill
\begin{description}[leftmargin=1.5em, labelindent=0pt, font=\normalfont\ttfamily, itemsep=8pt]

  \item[bio-phase-atlas]
  \normalfont
  Many biological systems are modelled by ordinary differential equations whose long-term behaviour (steady states, oscillations, switches) determines the biology. \texttt{bio-phase-atlas} takes such an ODE system, finds its equilibria and bifurcations, and uses the Z3 SMT solver to certify their existence and stability symbolically rather than numerically. For an outsider, it is the difference between ``my numerical integrator drifted toward this point'' and ``there is a proof that this point is an attractor''.

  \item[synbio-verifier]
  \normalfont
  Synthetic-biology designs (specified in standard formats like SBML/SBOL) often contain bugs that only show up after expensive lab work. \texttt{synbio-verifier} uses counterexample-guided abstraction refinement (CEGAR), the same loop used in modern hardware model checkers, to verify temporal-logic properties of genetic circuits, and to repair faulty designs by synthesising minimal edits. It scales to circuits with 50+ molecular species, which is roughly the size of recently published designs.

  \item[guideline-polypharmacy-verify]
  \normalfont
  Patients on several medications are typically governed by several disease-specific clinical guidelines that were never co-designed; combinations can recommend conflicting actions. \texttt{guideline-polypharmacy-verify} encodes guidelines as temporal contracts and pharmacokinetics as continuous-time semantics, and checks whether running them simultaneously can lead to dangerous regimens (e.g., overdose windows, contraindicated co-prescriptions). For a non-medical reader, it is contract-based verification of policy interactions, applied to medicine.

\end{description}

\item[Optimization]\hfill
\begin{description}[leftmargin=1.5em, labelindent=0pt, font=\normalfont\ttfamily, itemsep=8pt]

  \item[bilevel-compiler-intersection-cuts]
  \normalfont
  Bilevel programs (``the leader optimises subject to a follower also optimising'') model interactions between a regulator and a regulated agent, and are notoriously hard. \texttt{bilevel-compiler-intersection-cuts} compiles bilevel mixed-integer programs into single-level form and tightens the relaxation with intersection cuts---a geometric technique borrowed from mixed-integer programming. The deliverable is a black-box solver people can apply without learning the underlying cutting-plane theory.

  \item[bilevel-tight-reformulation]
  \normalfont
  A complementary approach to bilevel solving is the KKT reformulation, which replaces the inner optimisation by its first-order optimality conditions. The classical version is usually too loose to be practical. \texttt{bilevel-tight-reformulation} computes tight, problem-specific KKT reformulations and proves their correctness, turning theoretically clean but historically slow techniques into something that solves real instances.

  \item[pareto-reg-trajectory-synth]
  \normalfont
  AI systems deployed across jurisdictions face stacked regulatory regimes (EU AI Act, sector-specific rules, local data laws) whose obligations sometimes pull in opposite directions. \texttt{pareto-reg-trajectory-synth} formalises these as constraints over a trajectory of system behaviour and computes the Pareto frontier of compliance trade-offs. It does not pick the ``best'' answer; it shows decision-makers exactly which obligations they are trading against which, with a formal artefact behind each point.

\end{description}

\item[Security]\hfill
\begin{description}[leftmargin=1.5em, labelindent=0pt, font=\normalfont\ttfamily, itemsep=8pt]

  \item[zk-nlp-scoring]
  \normalfont
  When a vendor reports a benchmark score (BLEU, ROUGE, F1) for a closed model, you usually have to trust them. \texttt{zk-nlp-scoring} computes those metrics inside a STARK zero-knowledge proof: the verifier learns the score and verifies cryptographically that it was computed correctly on the claimed dataset, without seeing model outputs. For non-cryptography people: ``the leaderboard you can audit without the model owner having to publish their model''.

  \item[spectral-decomposition-oracle]
  \normalfont
  Eigenvalue decompositions sit inside countless numerical pipelines, but textbook libraries return only floating-point estimates. \texttt{spectral-decomposition-oracle} returns eigenvalues and eigenvectors together with rigorous, machine-checkable error bounds, so downstream code can branch on whether the precision is sufficient. To an outside reader: it adds an interval around every number a spectral library would normally return.

  \item[wasserstein-bounds]
  \normalfont
  Distribution-shift monitors watch streams of data and alert when the distribution drifts. Many of them give heuristic alarms with no statistical guarantees. \texttt{wasserstein-bounds} uses concentration inequalities to compute provably valid Wasserstein distance bounds that hold for any sample size, including small samples. The practical effect: a monitor whose false-alarm and miss rates can be controlled by a parameter rather than tuned by hand.

\end{description}

\item[NLP/audio]\hfill
\begin{description}[leftmargin=1.5em, labelindent=0pt, font=\normalfont\ttfamily, itemsep=8pt]

  \item[nlp-metamorphic-localizer]
  \normalfont
  Multi-stage NLP pipelines (tokenise, retrieve, classify, summarise) often have bugs that are hard to localise because each stage transforms the input. \texttt{nlp-metamorphic-localizer} pairs metamorphic testing---perturbing inputs in ways the output \emph{should} be invariant to---with causal fault localisation that pinpoints which stage's behaviour deviates from invariance. The result is not just ``the pipeline is buggy'' but ``stage 3's tokeniser is the cause''.

  \item[perceptual-sonification-compiler]
  \normalfont
  Sonification turns data streams into sound for monitoring. Most ad-hoc sonifications waste perceptual bandwidth (humans cannot tell two close pitches apart) or miss real-time deadlines. \texttt{perceptual-sonification-compiler} takes a high-level mapping from data dimensions to audio dimensions and compiles a bounded-latency renderer that maximises mutual information between data and percept, subject to psychoacoustic constraints. To a graphics person, it is the audio analogue of an information-preserving visual encoding.

  \item[sim-conservation-auditor]
  \normalfont
  Physics simulations are supposed to conserve quantities (energy, momentum, mass), but in practice their numerical integrators leak these by small amounts that accumulate over time. \texttt{sim-conservation-auditor} runs a simulation with certified numerical integrators and reports rigorous bounds on per-step conservation-law violations. For a graphics or robotics simulation user, it answers ``can I trust this simulation over a long horizon, and by how much?''~with a number rather than a vibe.

\end{description}

\item[Graphics/XR]\hfill
\begin{description}[leftmargin=1.5em, labelindent=0pt, font=\normalfont\ttfamily, itemsep=8pt]

  \item[choreo-xr-interaction-compiler]
  \normalfont
  Mixed-reality applications coordinate user movements, controller events, and virtual objects in time and space; bugs that strand a user (e.g., a menu blocking the only exit pose) are easy to ship and hard to find by playtesting. \texttt{choreo-xr-interaction-compiler} represents an interaction as a ``choreography'' (a temporal-spatial protocol) and uses CEGAR to verify safety properties such as deadlock-freedom and reachability of every intended state. For a graphics developer, it is a static analyser for the interaction layer that sits above the rendering loop.

  \item[xr-affordance-verifier]
  \normalfont
  Reach-based XR affordances assume an ``average'' user; many designs subtly exclude users with shorter arms, wheelchair seating, or different head-mounted-display fits. \texttt{xr-affordance-verifier} formally verifies that every interactable element is reachable across a parametric population of body shapes and seated/standing configurations, returning concrete failing configurations when it is not. To a graphics or HCI person, it is accessibility testing turned into a model-checking problem with a reach-set abstraction.

  \item[spatial-hash-compiler]
  \normalfont
  ``Given a point, which mesh face / volume cell / collision proxy contains it?''~appears everywhere in graphics, physics, and GIS. \texttt{spatial-hash-compiler} compiles these spatial queries into a data structure based on \emph{geometric perfect hashing}: $O(1)$ deterministic lookup, no chaining, no probing, with measured throughput above 600M queries/sec. The contribution is essentially a domain-specific hash function generator that exploits the geometry of the input rather than treating points as arbitrary keys.

  \item[tensor-train-modelcheck]
  \normalfont
  Stochastic models of chemical reactions blow up combinatorially: tracking $n$ species with $k$ molecules each gives $k^n$ states, well beyond what classical model checkers handle. \texttt{tensor-train-modelcheck} represents the joint distribution of species counts as a tensor train (a low-rank tensor decomposition popular in numerical linear algebra) and verifies continuous-stochastic-logic (CSL) properties on this compressed representation. For a non-formal-methods reader, the analogy is using the SVD to compress an image: the same trick, applied to the state space of a probabilistic system.

\end{description}

\end{description}
\endgroup

\end{document}